\documentclass[aps,prb,floatfix,superscriptaddress,amsmath,amssymb,noshowpacs,twocolumn]{revtex4-1}
\usepackage{graphics,graphicx,epsfig,amsmath,amssymb,epstopdf,wasysym,comment,esint}
\usepackage[T1]{fontenc}
\usepackage{frcursive}
\usepackage{color}
\newenvironment{frcseries}{\fontfamily{frc}\selectfont}{}

\usepackage{subfigure}
\allowdisplaybreaks

\newcommand{\be}{\begin{equation}}
\newcommand{\ee}{\end{equation}}
\newcommand{\bea}{\begin{eqnarray}}
\newcommand{\eea}{\end{eqnarray}}
\newcommand{\ba}{\begin{eqnarray*}}
\newcommand{\ea}{\end{eqnarray*}}
\newcommand{\dagga}{{\phantom{\dagger}}}

\newcommand{\bq}{\mathbf{q}}

\newcommand{\bk}{\mathbf{k}}

\newcommand{\bp}{\mathbf{p}}

\newcommand{\br}{\mathbf{r}}
\newcommand{\Ima}{{\text{Im}}}
\newcommand{\Rea}{{\text{Re}}}
\newcommand{\dis}{\displaystyle}

\newcommand{\fract}[2]{\frac{\dis \;#1\;}{\dis \;#2\;}}

\newcommand{\eqn}[1]{(\ref{#1})}

\newcommand{\ep}{{\epsilon}}

\newcommand{\bw}{\begin{widetext}}
\newcommand{\ew}{\end{widetext}}
\newenvironment{eqs}%
{\begin{equation} \begin{aligned}}%
{\end{aligned} \end{equation} }
\newcommand{\beal}{\begin{eqs}}
\newcommand{\eal}{\end{eqs}}

\newcommand{\bd}[1]{{\boldsymbol{#1}}}
\begin{document}

\title{Landau-Fermi liquids without quasiparticles}
\author{M.~Fabrizio}
\affiliation{International School for
  Advanced Studies (SISSA), Via Bonomea
  265, I-34136 Trieste, Italy} 

\begin{abstract}
Landau-Fermi liquid theory is conventionally believed to hold whenever the 
interacting single-particle density of states develops a $\delta$-like component 
at the Fermi surface, which is associated with quasiparticles. 
Here we show that a microscopic justification can be actually achieved under more general 
circumstances, even in case coherent quasiparticles are totally missing and the interacting 
single-particle density of states vanishes at the chemical potential as consequence of 
a pole singularity in the self-energy.   
\end{abstract}


\maketitle

\section{Introduction}
\label{SectI}

Landau's Fermi liquid theory~\cite{Landau1,Landau2} is a cornerstone of modern quantum many body 
physics, and represents by now a chief paradigm for interacting Fermi liquids at low temperatures, even strongly interacting ones as $^3$He~\cite{VollhardtRMP} and heavy fermions~\cite{Rice&Ueda-PRB1986}. \\
Its microscopic justification relies on the hypothesis~\cite{Nozieres&Luttinger-1,Nozieres&Luttinger-2} that the interacting single particle density of states (DOS), $\mathcal{A}(\ep,\bk)$, for energy $\ep\to 0$ 
and momentum approaching the Fermi surface, $\bk\to \bk_F$, defined through the Luttinger 
theorem~\cite{Luttinger&Ward,Luttinger}, becomes a $\delta$-function 
\beal
\mathcal{A}(\ep\to 0,\bk\to \bk_F) \simeq z_\bk\,\delta\big(\ep-\ep_*(\bk)\big)\,,\label{FL-assumption-1}
\eal
where the weight $z_\bk <1$ is the quasiparticle residue, which measures how much of a quasiparticle is contained in the physical single-particle excitation, and $\ep_*(\bk)$ the quasiparticle dispersion that vanishes at $\bk=\bk_F$.  
The validity of Eq.~\eqn{FL-assumption-1} can be verified order by order in perturbation theory, as we shall later 
discuss. \\

\noindent 
The Landau-Fermi liquid theory for a bulk of interacting fermions was phenomenologically extended by Nozi\`eres~\cite{Nozieres-localFL} to describe the Kondo regime of a quantum impurity model, what 
is commonly refereed to as a \textit{local} Landau-Fermi liquid, and later justified microscopically, see, e.g., Ref.~\onlinecite{Fred}. However, such generalisation to quantum impurities poses a puzzle that is the actual 
motivation of the present work, and which we now discuss through a specific example. \\
Let us consider the model of two Anderson impurities, each hybridised with its own bath, and coupled to each other by an antiferromagnetic exchange $J$~\cite{Jones&Varma,Lorenzo-PRB2004,Ferrero_2007}.
This model has a quantum critical point at $J=J_*$~\cite{Jones&VarmaPRB, Affleck&Ludwig,Affleck&Ludwig&Jones} that separates the phase at $J<J_*$, where each impurity is 
Kondo screened by its bath, from the phase at $J>J_*$, where the two impurities lock by means of  $J$ into a spin-singlet state, no more available to Kondo screening. Both phases at $J<J_*$ and $J>J_*$ 
are local Fermi-liquids in Nozi\`eres' sense. However, the Fermi liquid behaviour at $J>J_*$ emerges from 
a state characterised by the impurity DOS that vanishes quadratically approaching the chemical potential, $\mathcal{A}(\ep)\sim \ep^2$, i.e., without displaying the peculiar   Abrikosov-Suhl resonance, which reflects a diverging impurity self-energy, $\Sigma(\ep) \sim 1/\ep$ ~\cite{Lorenzo-PRB2004,Ferrero_2007}. 
Despite such singular behaviour, apparently at odds with a Fermi liquid,  one can still justify the latter 
microscopically~\cite{Lorenzo-PRB2004}, which raises the question whether it is possible to follow backward the path from bulk to local Fermi liquids 
with singular self-energies. Should that be indeed the case, it would imply that a microscopic justification 
of the Landau-Fermi liquid theory can be achieved with a less stringent requirement than Eq.~\eqn{FL-assumption-1}. \\
This is actually the main outcome of the present work, which is organised as follows. 
In Sect.~\ref{SectII} we briefly recall the microscopic justification of Eq.~\eqn{FL-assumption-1}, hence 
of the conventional derivation of Landau-Fermi liquid theory, which we rederive  in Sect.~\ref{SectIII} under a more general hypothesis, which includes Eq.~\eqn{FL-assumption-1} as a particular case. The results are exploited to obtain the Landau-Fermi liquid expressions of the dynamical susceptibilities in 
Sect.~\ref{SectIV}, which allows deriving a kinetic equation for the Wigner quasi-probability distribution of 
"quasiparticles" in Sect.~\ref{SectV}. Section~\ref{SectVI} is devoted to concluding remarks.

\section{Conventional Fermi liquid hypothesis}
\label{SectII}

Let us recall the general 
expression of the Green's function in Matsubara frequencies $i\ep = i\, \pi\,(2n+1)\,T$, with $n\in\mathbb{Z}$ and $T$ the temperature, 
\be
G(i\ep,\bk) = \fract{1}{i\ep -\ep_\bk -\Sigma(i\ep,\bk)}\,,\label{G}
\ee
where $\ep_\bk$ is the non-interacting dispersion relation measured relative to the chemical potential $\mu$, and $\Sigma(i\ep,\bk)$ the self-energy. Its continuation in the complex plane, $G(\zeta,\bk)$ 
with $\zeta\in\mathbb{C}$, is analytic everywhere but on the real axis, where it generally develops  
a branch cut 
\bea
&&G(\ep+i\eta,\bk) - G(\ep-i\eta,\bk) \equiv G_+(\ep,\bk)-G_-(\ep,\bk) \nonumber \\
&&\qquad \qquad  = -2i\,\Ima G_+(\ep,\bk) \equiv -2\pi\,i\,\mathcal{A}(\ep,\bk)\,,\label{def:branch cut}
\eea
with $\ep\in\mathbb{R}$, $\eta$ an infinitesimal positive real number, and $\mathcal{A}(\ep,\bk)\geq 0$ the single particle DOS at momentum $\bk$, satisfying 
\beal
\int d\ep \,\mathcal{A}(\ep,\bk) =1\,.\label{normalisation}
\eal
$G_+(\ep,\bk)$ and $G_-(\ep,\bk)$ in \eqn{def:branch cut} are, respectively, the retarded and advanced Green's functions. It is thus possible to write 
\beal
G(\zeta,\bk) &= \int d\omega\;\fract{\;\mathcal{A}(\omega,\bk)\;}
{\;\zeta - \omega\;}\;.\label{G.vs.A}
\eal
Similarly, the self-energy in the complex 
frequency plane, $\Sigma(\zeta,\bk)$, is also analytic but on the real axis. As before, 
for $\ep\in\mathbb{R}$, and $\eta>0$ infinitesimal, 
\be
\Sigma( \ep \pm i\eta,\bk) \equiv \Sigma_\pm(\ep,\bk) 
= \Rea\Sigma_+(\ep,\bk) \pm i\,\Ima\Sigma_+(\ep,\bk)\,,
\ee
define retarded, $\Sigma_+(\ep,\bk)$, and advanced, $\Sigma_-(\ep,\bk)$, components 
of the self-energy. It follows that 
\be
\mathcal{A}(\ep,\bk) = \fract{1}{\pi}\,
\fract{-\Ima \Sigma_+(\ep,\bk)}{
\big(\ep-\ep_\bk -\Rea\Sigma_+(\ep,\bk)\big)^2 
+ \Ima \Sigma_+(\ep,\bk)^2} \,,\label{DOS}
\ee
thus $\Ima\Sigma_+(\ep,\bk) \leq 0$. \\
In a conventional Fermi liquid, the Fermi surface (FS), $\bk=\bk_F$, is defined 
through~\cite{Luttinger}
\beal
\ep_{\bk_F} + \Rea\Sigma_+(0,\bk_F) =0\,,
\eal
while the quasiparticle dispersion by  
\beal
\ep_*(\bk)-\ep_\bk -\Rea\Sigma_+\big(\ep_*(\bk),\bk\big)=0\,,
\eal
so that, by definition, $\ep_*(\bk_F)=0$. The important observation is that, order by order in perturbation theory, the following result holds for $\bk$ close to the FS
\beal
-\Ima\Sigma_+(\ep\to 0,\bk) &= \Gamma(\bk)\,\ep^2 + \text{O}(\ep^4) \,.
\label{FL-assumption-2}
\eal
It follows that, expanding \eqn{DOS} for $\ep\simeq \ep_*(\bk)$, one finds 
\beal
\mathcal{A}(\ep,\bk) 
&\simeq \fract{Z_*\big(\ep_*(\bk),\bk\big)}{\pi}\\
&\qquad 
\fract{\gamma_*(\bk)\,\ep_*(\bk)^2}{
\Big(\ep-\ep_*(\bk) \Big)^2 
+ \gamma_*(\bk)^2\,\ep_*(\bk)^4} \\
&\underset{\bk\to\bk_F}{\xrightarrow{\hspace*{0.6cm}}} \;
Z_*\big(\ep_*(\bk),\bk\big)\,\delta\big(\ep-\ep_*(\bk)\big)\,,
\eal
thus Eq.~\eqn{FL-assumption-1} with $z_\bk = Z_*\big(\ep_*(\bk),\bk\big)$, where the formal definition of the quasiparticle residue reads 
\beal
Z_*(\ep,\bk) &\equiv \Bigg(1 - 
\fract{\partial \Rea\,\Sigma_+(\ep,\bk)}{\partial \ep}\Bigg)^{-1}\;,
\label{def:general-Z}
\eal 
and $\gamma_*(\bk)=Z_*\big(\ep_*(\bk),\bk\big)\,\Gamma(\bk)$. In conclusion, one can 
safely write for $\bk\sim \bk_F$
\be
\mathcal{A}(\ep,\bk) 
\simeq z_\bk\,\delta\big(\ep-\ep_*(\bk)\big) + 
\mathcal{A}_\text{inc}(\ep,\bk)\,, \label{FL-assumption-3}
\ee
with $\mathcal{A}_\text{inc}(\ep,\bk)$ a smooth function that carries the rest 
$1-z_\bk$ of the spectral weight, see Eq.~\eqn{normalisation}, 
and describes "incoherent" excitations as opposed to the "coherent" $\delta$-function component.  It follows, through 
\eqn{G.vs.A}, that 
\beal
G(i\ep,\bk) &\simeq \fract{Z_*\big(\ep_*(\bk),\bk\big)}{\;i\ep-\ep_*(\bk)\;} 
+ \int d\omega\;\fract{\mathcal{A}_\text{inc}(\omega,\bk)}{i\ep-\omega}\\
&\equiv G_\text{coh}(i\ep,\bk) + G_\text{inc}(i\ep,\bk)\,.
\label{FL-assumption-4}
\eal
We observe that $G_\text{coh}= Z_*\,G_0$, where $G_0(i\ep,\bk)$ is the Green's function 
of non-interacting electrons, the quasiparticles, with dispersion $\ep_*(\bk)$.  
Equation~\eqn{FL-assumption-3} coincides with the equation (2.15) of Ref.~\onlinecite{Nozieres&Luttinger-1}. Starting from that, we could retrace all steps of that work, as well as of the second of the series, Ref.~\onlinecite{Nozieres&Luttinger-2}, and thus 
recover microscopically the Landau-Fermi liquid theory. \\
However, the fact that each term in perturbation theory satisfies Eq.~\eqn{FL-assumption-2} 
does not guarantees that the sum of the perturbation series shares the same property.    

\section{Fermi liquid theory revised}
\label{SectIII}

Hereafter, we will reconsider the microscopic justification of Landau-Fermi liquid theory relaxing the hypothesis \eqn{FL-assumption-3}, or, equivalently, \eqn{FL-assumption-2}. For that, we shall have in mind a system of electrons, coupled to each other by a short range interaction~\cite{longrange}, with annihilation operators 
$c^\dagga_{a\bk}$, where $a$ includes all quantum numbers but momentum. The Green's function and the self-energy will be in general 
matrices in the $a$-space, or, if such basis is properly chosen, diagonal in $a$. In what follows, whenever not necessary, we discard the label $a$. Moreover, for further simplification, we shall not take into account the possible emergence of non trivial 
topological properties~\cite{Haldane-PRL2004,Son-PRL2012}, which has constituted one of the most notable extensions of Fermi liquids in recent years.  \\

\noindent
We shall here assume, in place of \eqn{FL-assumption-2}, that the 
following condition is satisfied:
\beal
&\lim_{\ep\to 0}\, Z_*(\ep,\bk)\;\Big(-\Ima\,\Sigma_+(\ep,\bk)\Big) \equiv  
\lim_{\ep\to 0}\,\gamma_*(\ep,\bk) \\
&\qquad = \lim_{\ep\to 0}\,\gamma_*(\bk)\,\ep^2 \to 0
\,,\label{gen-def:FL}
\eal
with $Z_*(\ep,\bk)$ defined by Eq.~\eqn{def:general-Z}. 
Equation~\eqn{gen-def:FL} is far less stringent than \eqn{FL-assumption-2}.
It is evidently satisfied if the conventional Fermi liquid 
hypothesis \eqn{FL-assumption-2} holds, but also in the extreme case of $\Sigma_+(\ep,\bk)$ singular at $\ep=0$, for instance,  
\beal
\Rea\,\Sigma_+(\ep,\bk) &\simeq \fract{\;\Delta(\bk)^2\;}{\ep}\;,\\ 
\Ima\,\Sigma_+(\ep,\bk) &\simeq -\pi\,\Delta(\bk)^2\,\delta(\ep) - \Gamma(\bk)\,,
\label{singular-Sigma}
\eal  
with $\Delta(\bk)\in\mathbb{R}$ and $\Gamma(\bk)\geq 0$, in which case the quasiparticle residue 
\beal
Z_*(\ep,\bk) &\simeq \fract{\ep^2}{\;\Delta(\bk)^2+\ep^2\;}\;\underset{\ep\to 0}{\longrightarrow}\; 0\;,
\eal
vanishes at the chemical potential, so do the particle DOS  
\be
\mathcal{A}(\ep,\bk) \simeq \fract{1}{\pi}\; \ep^2\;\fract{\Gamma(\bk)}{\;
\Delta(\bk)^4\;}\;\underset{\ep\to 0}{\longrightarrow}\;0\,.\label{p-DOS}
\ee
This is exactly the bulk counterpart of the two-impurity model behaviour for $J>J_*$~\cite{Lorenzo-PRB2004,Ferrero_2007} mentioned in the Introduction. We intentionally did not specify any precise $\bk$-dependence of $\Sigma(\ep,\bk)$ in \eqn{singular-Sigma}, 
so to maintain the discussion as general as possible.\\
For later use, we define the "quasiparticle" DOS through 
\beal
\mathcal{A}_\text{qp}(\ep,\bk)&\equiv \fract{\mathcal{A}(\ep,\bk)}{\;Z_*(\ep,\bk)\;}\;,
\label{def:A_qp}
\eal
and the "quasiparticle" group velocity as
\beal 
\bd{v}_*(\ep,\bk) 
&\equiv Z_*(\ep,\bk)\,\Bigg(
\fract{\partial \ep_\bk}{\partial \bk} 
+ \fract{\partial \Rea\,\Sigma_+(\ep,\bk)}{\partial \bk}\Bigg)\,.
\label{def:v_*}
\eal
We note that, when the conventional Fermi liquid hypothesis \eqn{FL-assumption-3} holds, 
then, for small $\ep$, 
$\mathcal{A}_\text{qp}(\ep,\bk)\simeq \delta\big(\ep-\ep_*(\bk)\big)$, thus describing a genuine coherent quasiparticle, and the on-shell group velocity $\bd{v}_*\big(\ep_*(\bk),\bk\big)=\partial\ep_*(\bk)/\partial\bk$. However, even in the singular case of Eq.~\eqn{singular-Sigma}, the "quasiparticle" DOS of Eq.~\eqn{def:A_qp} 
is finite at the chemical potential $\ep=0$, though not $\delta$-like, despite the particle DOS \eqn{p-DOS} vanishes.

\subsection{Preliminaries}

The standard derivation of Landau-Fermi liquid theory starts from considering 
the generic expression of a correlation function~\cite{Nozieres&Luttinger-1}
\be
Q(i\omega,\bq) =\fract{1}{V}\,\sum_\bk\, T\sum_{\ep}\;R(i\ep+i\omega,\bk+\bq;i\ep,\bk)\;F(i\ep)\,,
\label{def-Q}
\ee
with $V$ the number of sites, and 
\be
R(i\ep+i\omega,\bk+\bq;i\ep,\bk) = G(i\ep+i\omega,\bk+\bq)\,G(i\ep,\bk)\,,
\label{def-R-Q}
\ee
to be evaluated at low temperature, for small $\omega>0$ and $q=|\bq|$, and analysing the properties of the kernel $R= G\,G$ in the sense of a distribution in $\ep$. 
In the conventional case,  
each Green's function can be written as in Eq.~\eqn{FL-assumption-4}, thus 
\beal
R &= G_\text{coh}\,G_\text{coh} + \dots \equiv  G_\text{coh}\,G_\text{coh} 
+ R_\text{inc}\\
&\equiv \Delta +R_\text{inc} = Z_*^2\,G_0\,G_0 +R_\text{inc}\,, \label{R-simple}
\eal
where, we recall, $G_0$ is the Green's function of non-interacting particles  
with dispersion $\ep_*(\bk)$. It follows that the expression of $\Delta$ in the sense of a 
distribution can be readily obtained through the well known expression of the 
Lindhard function,
\ba
G_0\,G_0 &=& \fract{\delta_{\ep,0}}{T}\;
\fract{f\big(\ep_*(\bk)\big)-f\big(\ep_*(\bk+\bq)\big)}
{\;i\omega-\ep_*(\bk+\bq)+\ep_*(\bk)\;}\,,\label{G0-G0}
\ea
where $f(\ep)$ is the Fermi distribution function (compare, e.g., with Eq.~(2.23) in Ref.~\onlinecite{Nozieres&Luttinger-1}). The main property of $Z_*^2\,G_0\,G_0$, which 
is actually at the hearth of Landau-Fermi liquid theory, is the non-analytic behaviour 
in the origin $\omega=q=0$, unlike $R_\text{inc}$ that is assumed to be analytic.\\
If we replace condition \eqn{FL-assumption-2} 
with \eqn{gen-def:FL}, we cannot anymore use Eq.~\eqn{FL-assumption-4}, and thus 
Eq.~\eqn{G0-G0}, to determine the analytic properties of the kernel $R$. However, for that purpose, we can instead follow the derivation of the local Landau-Fermi liquid theory in quantum impurity 
models~\cite{Fred,Lorenzo-PRB2004}. We thus consider a contour in the complex frequency plane, $i\ep\to \zeta\in\mathbb{C}$, which runs clockwise at infinity. Assuming 
that the integrand vanishes faster than $1/\zeta$ at infinity, 
\ba
0 &=& \fract{1}{V}\,\sum_\bk\,\oint \fract{d\zeta}{2\pi i}\; f(\zeta)\,R(\zeta+i\omega,\bk+\bq;\zeta,\bk)\, F(\zeta) \\
&=& Q(i\omega,\bq) + \dots\,,
\ea
where the dots take into account the singularities of $R$ and $F$. In particular, $R$ has generically 
two horizontal branch cuts, the real axis $\zeta = \ep$ and the axis 
$\zeta = -i\omega + \ep$, with $\ep \in\mathbb{R}$, which merge into a single one, just the real axis, when $\omega=0$. The contribution of the horizontal strip 
$\text{Im}\zeta\in\; ]\!-\!\omega,0[$ may not be analytic at $\omega=q=0$: 
it trivially vanishes if $\omega\to 0$ first than $q\to 0$, so called $q$-limit, but it may not in the 
opposite $\omega$-limit. On the contrary, the contributions from $\text{Im}\,\zeta \geq 0$ 
and $\text{Im}\,\zeta \leq -\omega$ do not have any apparent reason of non analyticity. The strip contribution reads 
\bw
\beal
Q_{sing}(i\omega,\bq) &= \fract{1}{V}\,\sum_\bk\, T\sum_{\ep_0-\omega\leq\ep\leq-\ep_0}\;G(i\ep+i\omega,\bk+\bq)\,G(i\ep,\bk)\,F(i\ep)\\
&=  -\fract{1}{V}\!\sum_\bk \int \!\!\fract{d\ep}{2\pi i} f(\ep)\,
\bigg[G_+(\ep,\bk)\,G_-(\ep-i\omega,\bk)\,F(\ep-i\omega)-G_+(\ep+i\omega,\bk)\,G_-(\ep,\bk)\,F(\ep)\bigg]\,,
\eal
\ew
where $\ep_0 = \pi\,T$ is the lowest fermionic Matsubara frequency, which, after the analytic continuation $i\omega\to \omega+i\eta$, with $\eta>0$ infinitesimal,  becomes
\bw
\beal
Q_{sing}(\omega,\bq)
&=-\fract{1}{V}\!\sum_\bk \int \!\!\fract{d\ep}{2\pi i} f(\ep)\,
\bigg[G_+(\ep,\bk)\,G_-(\ep-\omega,\bk)\,F(\ep-\omega)-G_+(\ep+\omega,\bk)\,G_-(\ep,\bk)\,F(\ep)\bigg]\\
&=\fract{1}{V}\!\sum_\bk \int \!\!\fract{d\ep}{2\pi i} \Big(f(\ep)\,F(\ep) 
-f(\ep+\omega)\,F(\ep+\omega)\Big)\, G_+(\ep+\omega,\bk)\,G_-(\ep,\bk)
\\
&\simeq \fract{1}{V}\!\sum_\bk \int \!\!\fract{d\ep}{2\pi i}\,\bigg(\!\!\!-\!\fract{\partial f(\ep)}{\partial \ep}\bigg)\,\omega\,F(\ep)\;
G_+(\ep+\omega,\bk)G_-(\ep,\bk)\\
&=  \fract{1}{V}\!\sum_\bk \int \!\!\fract{d\ep}{2\pi i}\,\bigg(\!\!\!-\!\fract{\partial f(\ep)}{\partial \ep}\bigg)\,F(\ep)\; \bigg(G_-(\ep,\bk)-G_+(\ep+\omega,\bk+\bq)\bigg)\\
&\qquad\qquad \qquad
\fract{\omega}{\;\omega + i\eta -\big(\ep_{\bk+\bq}-\ep_\bk\big) 
-\Big(\Sigma_+(\ep+\omega,\bk+\bq)-\Sigma_-(\ep,\bk)\Big)\;}\;.\label{Qsing-vero}
\eal
\ew
For small $\omega$ and $q$, recalling that $\Rea\Sigma_-=\Rea\Sigma_+$ 
while $\Ima\Sigma_-=-\Ima\Sigma_+$, 
and through equations 
\eqn{def:general-Z}, \eqn{gen-def:FL} and \eqn{def:v_*}, we can write 
\bw
\beal
&\fract{\omega}{\;\omega + i\eta -\big(\ep_{\bk+\bq}-\ep_\bk\big) 
-\Big(\Sigma_+(\ep+\omega,\bk+\bq)-\Sigma_-(\ep,\bk)\Big)\;} \simeq Z_*(\ep,\bk)\;\fract{\omega}{\;\omega + i\eta -\bd{v}_*(\ep,\bk)\cdot\bq + 2i\,\gamma_*(\ep,\bk)\;}\;. 
\label{denom}
\eal
\ew
Since the derivative of the Fermi distribution function in \eqn{Qsing-vero} implies that $\ep\sim T \sim 0$, 
if we assume Eq.~\eqn{gen-def:FL} valid, we can safely neglect $\gamma_*(\ep,\bk)$ in \eqn{denom} if either $\omega$ or $\bd{v}_*(T,\bk)\cdot\bq $ are much greater 
than $\gamma_*(\bk)\,T^2$. 
Coming back to \eqn{Qsing-vero}, and noting that 
\ba
G_-(\ep,\bk)-G_+(\ep+\omega,\bk+\bq) \simeq 2\pi\, i\,\mathcal{A}(\ep,\bk) + O(q,\omega)\,,
\ea
we can finally write 
\be
Q_{sing}(\omega,\bq)
=  \fract{1}{V}\!\sum_\bk \!\int \!\!d\ep\, \widetilde{\Delta}(\ep+\omega,\bk+\bq;\ep,\bk)\;
F(\ep)
\;,\label{Qsing-vero-1}
\ee
having defined the distribution kernel 
\bea
\widetilde{\Delta}(\ep+\omega,\bk+\bq;\ep,\bk)
&=&-\fract{\partial f(\ep)}{\partial \ep}\,
\mathcal{A}_\text{qp}(\ep,\bk)\,Z_*(\ep,\bk)^2\nonumber\\
&&\qquad \fract{\omega}{\;\omega -\bd{v}_*(\ep,\bk)\cdot\bq \;}\;,\label{Delta-tilde}
\eea
which is indeed non analytic at $\omega=q=0$, where $\mathcal{A}_\text{qp}(\ep,\bk)$ is 
defined by \eqn{def:A_qp}. In other words, the non analytic behaviour persists even if 
$R$ in Eq.~\eqn{def-Q} cannot be written as in \eqn{R-simple} in terms of non-interacting 
Green's functions, provided Eq.~\eqn{gen-def:FL} holds. 


Going back to Eq.~\eqn{def-Q}, we end up with the following expression
\beal
R &\equiv \widetilde{\Delta} + \widetilde{R}_\text{inc}\,,
\eal
where $\widetilde{R}_\text{inc}$ is analytic at the origin, and all non-analyticities are hidden 
in $\widetilde{\Delta}$. Specifically,  
\beal
\lim_{\omega\to 0}\,\lim_{q\to 0}\,\widetilde{\Delta}(i\ep+i\omega,\bk+\bq;i\ep,\bk) &\equiv \widetilde{\Delta}^\omega(i\ep,\bk) \not=0\,,\\
\lim_{q\to 0}\,\lim_{\omega\to 0}\,\widetilde{\Delta}(i\ep+i\omega,\bk+\bq;i\ep,\bk) &\equiv \widetilde{\Delta}^q(i\ep,\bk)  =0\,. 
\eal
We further define 
\beal
\Delta(i\ep+i\omega,\bk+\bq;i\ep,\bk)&\equiv 
\widetilde{\Delta}(i\ep+i\omega,\bk+\bq;i\ep,\bk)\\
&\qquad - \widetilde{\Delta}^\omega(i\ep,\bk)\,, 
\eal
whose expression on the real axis is 
\bea
\Delta(\ep+\omega,\bk+\bq;\ep,\bk)
&=&-\fract{\partial f(\ep)}{\partial \ep}\,
\mathcal{A}_\text{qp}(\ep,\bk)\,Z_*(\ep,\bk)^2\nonumber \\
&&\qquad \fract{\bd{v}_*(\ep,\bk)\cdot\bq}{\;\omega -\bd{v}_*(\ep,\bk)\cdot\bq \;}\;,
\label{def:Delta}
\eea
where now 
\beal
\lim_{\omega\to 0}\,\lim_{q\to 0}\,\Delta(i\ep+i\omega,\bk+\bq;i\ep,\bk) &\equiv \Delta^\omega(i\ep,\bk) =0\,,\\
\lim_{q\to 0}\,\lim_{\omega\to 0}\, \Delta(i\ep+i\omega,\bk+\bq;i\ep,\bk) &\equiv \Delta ^q(i\ep,\bk)  \\
=- \widetilde{\Delta}^\omega(i\ep,\bk) &\not= 0\,,
\eal
and, consequently, 
\beal
R &\equiv \Delta + R_\text{inc}\,,& 
R_\text{inc} &= \widetilde{R}_\text{inc} + \widetilde{\Delta}^\omega\,.
\label{def:R-usual}
\eal
The quantities $\Delta$ and $\widetilde\Delta$ coincide, respectively, 
with those in equations (2.23) and (2.33) of Ref.~\onlinecite{Nozieres&Luttinger-1}. 
Therefore, from this point on, we can simply follows all steps of Ref.~\onlinecite{Nozieres&Luttinger-1}, which we shall not repeat, and jump directly to 
the final results in the following sections.~\cite{nota-Gamma}

\section{Dynamic susceptibilities}
\label{SectIV}
Suppose the interacting Hamiltonian admits a conserved quantity $Q$. Then, in the basis in which the corresponding single-particle operator is diagonal, i.e., 
\beal
Q = \int d\br\,\rho_Q(\br) \equiv \sum_a\, \sum_\bk\, q_a(\bk)\, c^\dagger_{\bk a}\,c^\dagga_{\bk a}\,,
\eal
with $\rho_Q(\br)$ the density operator corresponding to $Q$, the Green's function is diagonal, too. For simplicity, we shall assume that the Green's function is actually independent of $a$. A smoothly varying external field $h_Q(t,\br)$, with Fourier component 
$h_Q(\omega,\bq)$, is coupled to $\rho_Q(\br)$, adding to the Hamiltonian the 
time-dependent perturbation $\delta H(t) = \int d\br\,h_Q(t,\br)\,\rho_Q(\br)$.  
Following Ref.~\onlinecite{Nozieres&Luttinger-1}, one can demonstrate that, at linear order in the field, the variation of the expectation value of $\rho_Q(\bq)$ reads 
\beal
\delta \langle\,\rho_Q(\bq)\,\rangle 
&= \chi_Q(\omega,\bq)\,h_Q(\omega,\bq)\,,
\eal
where the linear response function is given by 
\bw 
\beal
\chi_Q(\omega,\bq) &= -\fract{1}{V}\,\sum_\bk\, \int d\ep\, \bigg(\!\!-\fract{\partial f(\ep)}{\partial \ep}\bigg)\,
 \mathcal{A}_\text{qp}(\ep,\bk)\;
\fract{\bd{v}_*(\ep,\bk)\cdot\bq}{\;\omega - \bd{v}_*(\ep,\bk)\cdot\bq+i\eta\;}\;  \\
&\quad - \fract{1}{V^2}\,\sum_{\bk\bk'}\, 
\int d\ep\,d\ep'\; \bigg(\!\!-\fract{\partial f(\ep)}{\partial \ep}\bigg)\,\bigg(\!\!-\fract{\partial f(\ep')}{\partial \ep'}\bigg)\,\mathcal{A}_\text{qp}(\ep,\bk)\;
\fract{\bd{v}_*(\ep,\bk)\cdot\bq}{\;\omega - \bd{v}_*(\ep,\bk)\cdot\bq+i\eta\;}\\
&\qquad \qquad \qquad 
\sum_{aa'}\,q_a(\bk)\,q_{a'}(\bk')\,\text{A}_{a,a';a',a}(\ep\,\bk,\ep'\,\bk';\omega\,\bq)\,\mathcal{A}_\text{qp}(\ep',\bk')\;
\fract{\bd{v}_{*}(\ep',\bk')\cdot\bq}{\;\omega - \bd{v}_{*}(\ep',\bk')\cdot\bq+i\eta\;}\;,
\label{chi_Q}
\eal
\ew
assuming $\sum_a\,q_a(\bk)^2=1$ the normalisation of the conserved quantity, and having defined 
the "quasiparticle" scattering amplitudes 
\beal
&\text{A}_{a,b;b,a}(\ep\,\bk,\ep'\,\bp;\omega\,\bq) = Z_*(\ep,\bk)\,Z_*(\ep',\bp)\\
 &\qquad\;\Gamma_{a,b;b,a}(\ep+i\omega\,\bk+\bq,\ep'\,\bp;\ep'+i\omega\,\bp+\bq,\ep\,\bk)\,,
 \label{A.vs.Gamma}
\eal
where $\Gamma$ is the reducible vertex, with $q$ and $\omega$ limits 
$\text{A}^q_{a,b;b,a}(\ep\,\bk,\ep'\,\bp)$ and $\text{A}^\omega_{a,b;b,a}(\ep\,\bk,\ep'\,\bp)$, 
respectively. 
  \\
The thermodynamic susceptibility $\kappa_Q$ is related to the $q$-limit of the dynamical one 
$\chi_Q$, specifically,  
\ba
\kappa_Q &=& \chi^q_Q = \fract{1}{V}\sum_\bk\, \int d\ep\, \bigg(\!\!-\fract{\partial f(\ep)}{\partial \ep}\bigg)\;
 \mathcal{A}_\text{qp}(\ep,\bk)\;  \\
&&\; - \fract{1}{V^2}\,\sum_{\bk\bp}\, 
\int d\ep\,d\ep'\; \bigg(\!\!-\fract{\partial f(\ep)}{\partial \ep}\bigg)\,\bigg(\!\!-\fract{\partial f(\ep')}{\partial \ep'}\bigg)\\
&&\qquad \qquad\qquad \mathcal{A}_\text{qp}(\ep,\bk)\;\mathcal{A}_\text{qp}(\ep',\bp)\\
&& \qquad\qquad\quad \sum_{ab}\,q_a(\bk)\,q_b(\bp)\,\text{A}^q_{a,b;b,a}(\ep\,\bk,\ep'\,\bp)\;.
\ea
Since the local quasiparticle DOS is, by definition, 
\beal
\mathcal{A}_\text{qp}(\ep) 
\equiv \fract{1}{V}\sum_\bk\,\mathcal{A}_\text{qp}(\ep,\bk)
\;,
\eal
upon defining the Landau $\text{A}_Q$ parameter in channel $Q$ as  
\beal
&\text{A}_{Q}\; \mathcal{A}_\text{qp}(0)
\equiv \fract{1}{V^2}\,\sum_{\bk\bp}\,\sum_{ab}\,q_a(\bk)\,q_b(\bp)\\
&\qquad\qquad \int d\ep\,d\ep'\; \bigg(\!\!-\fract{\partial f(\ep)}{\partial \ep}\bigg)\,\bigg(\!\!-\fract{\partial f(\ep')}{\partial \ep'}\bigg)\\
&\qquad\qquad\quad \mathcal{A}_\text{qp}(\ep,\bk)\;\mathcal{A}_\text{qp}(\ep',\bp)\;\text{A}^q_{a,b;b,a}(\ep\,\bk,\ep'\,\bp)\,.
\eal 
we finally obtain, at zero temperature,   
\beal
\kappa_Q &= \mathcal{A}_\text{qp}(0)\,\Big( 1 - A_Q\Big)\,,\label{kappa-Q}
\eal
which is the standard Landau-Fermi liquid expression, but derived under the more general assumption \eqn{gen-def:FL}. In fact, the expression \eqn{kappa-Q} holds  
also in the case Eq.~\eqn{singular-Sigma} of a singular self-energy. yielding vanishing 
quasiparticle residue and particle DOS at the chemical potential. Nonetheless,  
the "quasiparticle" DOS is finite so as the zero temperature thermodynamic susceptibility. We also remark that a finite Landau $A_Q$ parameter in spite of a vanishing quasiparticle residue implies, through Eq.~\eqn{A.vs.Gamma}, that the reducible vertex $\Gamma$ is singular at the chemical potential. \\
We emphasise that the rather simple expression \eqn{chi_Q} of the linear response functions, which looks like that of weakly interacting (quasi)particles, holds only for density operators that refer to conserved quantities, for which one can use the Ward-Takahashi identity. Otherwise, the response functions contain additional 
observable-dependent parameters, see Eq.~(3.9) in Ref.~\onlinecite{Nozieres&Luttinger-1}; 
specifically, an additional constant that corresponds to the $\omega$-limit of the response function, vanishing for conserved quantities, and the $\omega$-limit of vertex corrections. 
The meaning of such difference is that only for conserved quantities we are guaranteed that 
the matrix element coupling the external field to the density of physical particles 
is the same as that one coupling to the density of quasiparticles, while for generic observables this ought not to be the case. \\

\section{Landau-Boltzmann equation}
\label{SectV}

The expression \eqn{chi_Q} of the linear response functions allows easily deriving 
a corresponding Boltzmann kinetic equation, which we believe worth showing explicitly. \\
We first associate to the expression \eqn{def:Delta} of $\Delta$ for real frequencies   
the components of a matrix $\hat{K}(\omega,\bq)$ in frequency, momentum, and quantum 
number $a$ space, through 
\beal
K_{\ep\bk a,\ep'\bk' a'}(\omega,\bq) &= \delta(\ep-\ep')\,\delta_{\bk,\bk'}\,
\delta_{a,a'}\\
& \qquad\qquad \Delta(\ep+\omega,\bk+\bq;\ep,\bk)\,.
\eal
Next we formally write
\beal
\delta\langle\rho_Q(\bq)\rangle &= \chi_Q(\omega\,\bq)\,h_Q(\omega,\bq)\\
&\equiv \fract{1}{V}\,\sum_{a\bk}\, \int d\ep\,q_a(\bk) \, 
\delta n_{\ep\bk a}(\omega,\bq)\,,
\eal
where $\delta n_{\ep\bk a}(\omega,\bq)$ are the components of the vector $\delta\bd{n}(\omega,\bq)$, which, through 
Eq.~\eqn{chi_Q}, satisfies  
\beal
\delta\bd{n} &= -\hat{K}\,\Big[\;1 + \hat{A}\, \hat{K}\Big]\,\bd{V}\,,
\eal
or, equivalently, 
\beal
\Big[1 + \hat{A}\,\hat{K}\Big]^{-1}\,\hat{K}^{-1}\,\delta\bd{n} &= 
-\bd{V} \,,\label{LB:def-1}
\eal
having defined $\hat A$ the matrix with elements $
A_{\ep\bk a,\ep'\bk' a'}(\omega,\bq)=\text{A}_{a,a';a',a}(\ep\,\bk,\ep'\,\bk';\omega\,\bq)$,  
and $\bd{V}$ the vector with components $V_{\ep \bk a}(\omega,\bq) = q_a(\bk)\,h_Q(\omega,\bq)$. We next introduce the standard Landau's $f$ parameters through 
\beal
\Big[ 1 + \hat{A}\,\hat{K}\Big]^{-1} &\equiv \Big[ 1 - \hat{f}\,\hat{K}\,\Big]\,,
\label{A.vs.f}
\eal
so that Eq.~\eqn{LB:def-1} becomes
\beal
\Big[ 1 - \hat{f}\,\hat{K}\,\Big]\,\hat{K}^{-1}\,\delta\bd{n} &= 
\hat{K}^{-1}\,\delta\bd{n} -\hat{f}\,\delta\bd{n} =
-\bd{V}\,.\label{LB:def-2}
\eal
Multiplying both sides of Eq.~\eqn{LB:def-2} by $\big(\omega-\bd{v}_*(\ep,\bk)\cdot\bq\big)\,\hat{K}$ we finally obtain the equation 
\bw 
\beal
0 &= \Big(\omega - \bd{v}_*(\ep,\bk)\cdot\bq\Big)\,\delta n_{\ep \bk a}(\omega,\bq)\\
&\qquad  + \fract{\partial f(\ep)}{\partial \ep}\;
\mathcal{A}_\text{qp}(\ep,\bk)\;
\bd{v}_*(\ep,\bk)\cdot\bq\;
\Bigg\{\fract{1}{V}\sum_{\bk' a'}\int d\ep'
f_{\ep\bk a, \ep'\bk' a'}\; \delta n_{\ep'\bk' a'}(\omega,\bq) - q_a(\bk) \,h_Q(\omega,\bq)\,\Bigg\}\\
&\equiv \omega \delta n_{\ep \bk a}(\omega,\bq)
- \bd{v}_*(\ep,\bk)\cdot\bq\,\delta \overline{n}_{\ep \bk a}(\omega,\bq)
-q_a(\bk)\,\fract{\partial f(\ep)}{\partial \ep}\;
\mathcal{A}_\text{qp}(\ep,\bk)\;
\bd{v}_*(\ep,\bk)\cdot\bq\; \,h_Q(\omega,\bq)
\,,
\eal
\ew
where we assumed that the dependence of $f$ upon $\omega$ and $\bq$ is
negligible when they are both small, and, by definition, 
\beal
\delta\overline{\bd{n}} &= \Big[\,1-\hat{K}^q\,\hat{f}\,\Big]\,\delta\bd{n}\,.
\label{local-equilibrium}
\eal
It follows that the inverse Fourier transform $\delta n_{\ep \bk a}(t,\br)$ satisfies 
\bea
0&=&\delta \dot{n}_{\ep \bk a}(t,\br)
+ \bd{v}_*(\ep,\bk)\cdot\bd{\nabla}\delta \overline{n}_{\ep \bk a}(t,\br)
\label{transport-eq}\\
&&\quad + q_a(\bk)\;\fract{\partial f(\ep)}{\partial \ep}\,
\mathcal{A}_\text{qp}(\ep,\bk)\,
\bd{v}_*(\ep,\bk)\cdot\bd{\nabla}h_Q(t,\br)\,,
\nonumber
\eea
which can be interpreted as the standard Landau-Boltzmann kinetic equation once we identify 
$\delta n_{\ep \bk a}(t,\br)$ and $\delta \overline{n}_{\ep \bk a}(t,\br)$, respectively,  with the deviations from global and local equilibrium,  see chapter 1 in Ref.~\onlinecite{Pines&Nozieres}, 
of the Wigner quasi probability distribution of quasiparticles. \\
In addition, the interpretation in terms of a semiclassical kinetic equation requires that the group velocity defined in Eq.~\eqn{def:v_*} is equivalent to 
\beal
\bd{v}_*(\ep,\bk) &\equiv \fract{\partial \,\ep_*(\ep,\bk)}{\partial \bk}\;,\label{e.vs.v}
\eal
where $\ep_*(\ep,\bk)$ must be identified with the quasiparticle energy, 
and that the derivative with respect to $\bk$ of the Wigner distribution at equilibrium must 
correspond to 
\beal
\fract{\partial n^0_{\ep\bk a}}{\partial \bk} &= \fract{\partial f(\ep)}{\partial \ep}\;\mathcal{A}_\text{qp}(\ep,\bk)\;\bd{v}_*(\ep,\bk)\,. \label{dn.vs.k}
\eal
We shall explicitly prove the last equality in the Appendix.
Through Eqs.~\eqn{e.vs.v} and \eqn{dn.vs.k} we can therefore rewrite Eq.~\eqn{transport-eq} as 
\beal
0&= \delta \dot{n}_{\ep \bk a}(t,\br) 
+ \fract{\partial \,\ep_*(\ep,\bk)}{\partial \bk}\cdot\bd{\nabla}\delta \overline{n}_{\ep \bk a}(t,\br)
\\
& \qquad + q_a(\bk)\,\fract{\partial n^0_{\ep \bk a}}{\partial \bk}\!\cdot\!\bd{\nabla}h_Q(t,\br)\,,
\label{transport-eq-final}
\eal
which has now truly the form of  the conventional Landau-Boltzmann kinetic equation, and entails 
a Landau's energy functional in absence of the external field 
\bea
F\big[\delta n\big] &=& \sum_{\bk a}\int \!\! d\br \,d\ep\, \bigg\{\;
\ep_*(\ep,\bk) \;\delta n_{\ep \bk a}(t,\br)\nonumber \\
&& \qquad + \fract{1}{2V}\sum_{\bk' a'}\int  d\ep'\,
f_{aa'}(\ep\,\bk, \ep'\,\bk')\\
&& \qquad\qquad \qquad \qquad
\delta n_{\ep \bk a}(t,\br)\,\delta n_{\ep' \bk' a'}(t,\br)
\,\bigg\}\,.\nonumber
\eea\\

\noindent
We end remarking that for a conventional Fermi liquid, where 
$\mathcal{A}_\text{qp}(\ep,\bk) = \delta\big(\ep-\ep_*(\bk)\big)$, one can readily integrate 
over $\ep$ both sides of Eq.~\eqn{transport-eq-final} and recover the standard Landau-Boltzmann kinetic equation for the integrated 
$\delta n_{\bk a}(t,\br) = \int d\ep\, \delta n_{\ep \bk a}(t,\br)$. However, Eq.~\eqn{transport-eq-final} remain valid also when $\mathcal{A}_\text{qp}(\ep,\bk) \not= \delta\big(\ep-\ep_*(\bk)\big)$, in which case the dependence 
of the quasiparticle DOS and group velocity $\bd{v}_*(\ep,\bk)$  
on the frequency $\ep$, which may also be rather non trivial, must be explicitly taken into account.    

\section{Conclusions}
\label{SectVI}
We have shown that the Landau-Fermi liquid low-temperature expressions of the dynamical 
susceptibilities in the long wavelength limit and small frequency, as well as the corresponding Boltzmann kinetic equation, can be microscopically justified even if the interacting single-particle Green's function does not have a quasiparticle pole near the 
chemical potential. \\
This result may not come as a surprise. For instance, also one dimensional Luttinger liquids~\cite{D&L,Solyom,Haldane-LL-1981,Glazman-RMP2012}, despite not fulfilling Eq.~\eqn{FL-assumption-1}, have dynamical susceptibilities similar to Fermi liquids in the long wavelength and low frequency limit. Specifically, in Luttinger liquids such behaviour arises as a consequence of an emerging symmetry that ensures, asymptotically, the independent conservation of electron densities at the two different Fermi points, which could be a 
mere one dimensional feature, or hide a more fundamental link between Luttinger and 
Fermi liquids~\cite{haldane2005luttingers}. \\
What is remarkable of our results is that a Fermi liquid behaviour emerges even in the worst case of a self-energy with a pole singularity at the chemical potential, which might look 
the furthest possible from a conventional Landau-Fermi liquid. We did not consider explicitly 
any model self-energy, but the extreme case of Eq.~\eqn{singular-Sigma}, where 
the main assumption \eqn{gen-def:FL} is verified, and thus a Landau-Fermi liquid 
description holds. However, given the generality of that assumption, it is well possible that purported non-Fermi liquid properties sometimes observed in correlated materials might 
be actually reconciled with the broader Fermi liquid scenario we have here uncovered.

\section*{Acknowledgments}
\noindent
I am extremely grateful to Claudio Castellani for helpful discussions and comments. 
This work has received funding from the European Research Council (ERC) under the European Union's Horizon 2020 research and innovation programme, Grant agreement No. 692670
``FIRSTORM''.

\appendix

\section{Luttinger theorem and quasiparticle equilibrium distribution}
\label{appendix}

While previously we defined the energies $\ep_\bk$ relative to the chemical potential $\mu$,   
in this appendix we move back to absolute units, so that the Green's function
\beal
G(i\ep,\bk) &\to \fract{1}{i\ep-\ep_\bk+\mu-\Sigma(i\ep,\bk)}\;,
\eal
depends on $\mu$, as well as the self-energy does, though we shall not indicate such explicit dependence.\\    
According to the Luttinger theorem~\cite{Luttinger} the (conserved) number of $a$-particles 
per site can be written as 
\beal
\rho_a &= \fract{1}{V}\sum_{\bk}\int \fract{d\ep}{\pi}f(\ep)\,\fract{\partial \delta_a(\ep,\bk)}{\partial \ep}\,,\label{Luttinger-sum}
\eal
where, dropping the label $a$ whenever not needed, 
\bea
\delta(\ep,\bk) &=& \pi + \Ima\ln G_{+}(\ep,\bk)\nonumber\\
&=& \tan^{-1}\fract{-\Ima G_+(\ep,\bk)}{-\Rea G_+(\ep,\bk)}
\,,\label{MB-phase-shift}
\eea
is the many-body phase shift. By definition, $\delta(\ep,\bk)\in [0,\pi]$, 
and vanishes at $\ep\to-\infty$, while reaches $\pi$ at $\ep\to\infty$, consistently with 
each momentum state accommodating at most a single electron species. We remark 
that $\delta(\ep,\bk)$ is in general not monotonous, and may jump back and forth between 
0 and $\pi$. 
The derivative of Eq.~\eqn{Luttinger-sum} with respect to $\mu_a$ corresponds to the thermodynamic compressibility of the species $a$, and reads   
\beal
\kappa_a &\equiv \fract{\partial \rho_a}{\partial\mu_a} =  \fract{1}{V}\sum_{\bk}\int \fract{d\ep}{\pi} f(\ep)\,
\fract{\partial^2 \delta_a(\ep,\bk)}{\partial \ep\,\partial \mu_a}\;.\label{compre}
\eal
Thermodynamic stability requires $\kappa_a \geq 0$. 
We may state that the $a$ electron species is metallic if $\kappa_a>0$, while is insulating
if $\kappa_a=0$. Note that at low temperatures the integral involves energies within 
a small window of order $T$ around $\ep=0$. 
Let us discuss the behaviour of $\delta(\ep,\bk)$ at vanishing temperatures in different cases.

\subsection{Systems with a single-particle gap}
Suppose that the single-particle DOS, 
$\mathcal{A}(\ep,\bk)=-\pi\,\Ima G_+(\ep,\bk)$, vanishes for $\ep$ in a whole interval $X_\text{ins}(\bk)$, $\forall\,\bk$, which includes $\ep=0$ and is definitely much 
wider than the temperature.
By the Kramers-Kr\"{o}nig relations it follows that $-\Rea G_+(\ep,\bk)$ must cross 
zero with positive slope for $\ep = \ep_\text{root}(\bk) -\mu\in X_\text{ins}(\bk)$. 
Correspondingly, the phase shift for $\ep\in X_\text{ins}(\bk)$ reads
\beal
\delta(\ep,\bk) &= \pi\,\theta\big(\ep_\text{root}(\bk)-\mu-\ep\big)\,.
\eal
We can envisage two different cases. If $\ep_\text{root}(\bk)$ is not pinned at the 
chemical potential, Eq.~\eqn{Luttinger-sum} at zero temperature simplifies into 
\beal
\rho = \fract{1}{V}\,\sum_\bk\, \theta\big(\ep_\text{root}(\bk)-\mu\big)\,,
\eal
which implies that the total density 
corresponds to the volume that contains all $\bk$ such that 
$\ep_\text{root}(\bk) > \mu$,
and thus enclosed by the Luttinger surface (LS) defined through $\ep_\text{root}(\bk) = \mu$, 
or, equivalently, 
\beal 
\Rea G_+(0,\bk)=0 \qquad \forall\,\bk \in \text{LS}\,.\label{def:LS}
\eal
This is, e.g., the case of 
a BCS superconductor, where $\ep_\text{root}(\bk) = -\ep_{-\bk}+2\mu=-\ep_\bk+2\mu$, so that  the LS is just 
the non-interacting Fermi surface, and the compressibility \eqn{compre} is equal to the non-interacting one. \\
It may instead happen that $\ep_\text{root}(\bk)$ is pinned at the chemical potential, i.e.,  $\ep_\text{root}(\bk)=\mu$, $\forall\,\bk$, so that $\delta(\ep,\bk)$ 
jumps from $\pi$ to 0 right at $\ep=0$. In this case, Eq.~\eqn{Luttinger-sum} becomes, at zero temperature, 
\beal
\rho_a &= \fract{1}{V}\sum_{\bk}\int \fract{d\ep}{\pi}f(\ep-\mu)\,\fract{\partial \delta_a(\ep,\bk)}{\partial \ep}\\
&= -f(0) + \fract{1}{V}\sum_{\bk}\fint \fract{d\ep}{\pi}f(\ep-\mu)\,\fract{\partial \delta_a(\ep,\bk)}{\partial \ep}\\
&= - f(0) + \fract{1}{V}\,\sum_{\bk}\,\fract{\delta_a(\mu^-,\bk)}{\pi} = \fract{1}{2}\,,
\eal 
where $\fint \dots$ is the Cauchy principal value of the integral, which implies that the state $a$ is half-filled. Moreover, the compressibility \eqn{compre} vanishes, as exprected for an insulator. Such circumstance in which $\ep_\text{root}(\bk)=\mu$ defines, e.g., 
a Mott insulator, and entails a self energy with a pole at $\ep=0$. 

\subsection{Systems with gapless single-particle excitations satisfying Eq.~\eqn{gen-def:FL}}
Gapless single-particle excitations correspond to a DOS $\mathcal{A}(\ep,\bk)$ 
smooth and finite in a finite interval around $\ep=0$ and for momenta $\bk$ within regions of the Brillouin zone with non-zero measure. We further assume the validity of Eq.~\eqn{gen-def:FL}, which allowed us recovering the Landau-Fermi liquid theory. In this case it is 
straightforward to show~\cite{Nozieres&Luttinger-1} that the compressibility as defined in \eqn{compre} has the same expression as that of Eq.~\eqn{kappa-Q}, and thus is finite as expected for a metallic state. It is therefore tempting to make the association 
\beal
n^0_{\ep\bk a} &\overset{?}{\equiv} f(\ep)\;\fract{1}{\pi}\;\fract{\partial \delta_a(\ep,\bk)}{\partial \ep}\;,
\label{n.vs.phase-shift}
\eal
between the equilibrium distribution of "quasiparticles" and the derivative of the many-body 
phase shift. However, such equivalence, though building a suggestive link to quantum impurity models~\cite{nota-phase-shift}, is dubious, since the right hand side of 
Eq.~\eqn{n.vs.phase-shift} may be negative or even singular, as it is the case for the 
self-energy in Eq.~\eqn{singular-Sigma}. However, the Landau hypothesis of adiabatic evolution~\cite{Landau1} only refers to low energy excitation, with no reference to the ground state. In other words, what really matters is the variation of the "quasiparticles" distribution with respect to the equilibrium one, the latter playing no role in the theory. 
Therefore, the most correct association is not Eq.~\eqn{n.vs.phase-shift} but rather 
\beal
\delta\Big(n^0_{\ep\bk a}\Big) &\equiv \delta\bigg(\,f(\ep)\;\fract{1}{\pi}\;\fract{\partial \delta_a(\ep,\bk)}{\partial \ep}\,\bigg)\\
&= \delta\bigg(\,-\fract{\partial f(\ep)}{\partial \ep}\;\fract{\delta_a(\ep,\bk)}{\pi}\,\bigg)\;,
\label{dn.vs.phase-shift}
\eal  
where $\delta\big(\dots\big)$ denotes the variation with respect to internal or thermodynamic variables, and the last expression on the right hand side is obtained after integration by part of Eq.~\eqn{Luttinger-sum}. \\
Indeed, by the definition of $\delta(\ep,\bk)$ in Eq.~\eqn{MB-phase-shift} it is 
straightforward to show that 
\beal
\fract{\partial n^0_{\ep\bk a}}{\partial \bk} &= \fract{1}{\pi}\,\bigg(\!\!-\!
\fract{\partial f(\ep)}{\partial \ep}\!\bigg)\, \fract{\partial \delta_a(\ep,\bk)}{\partial \bk}\,,
\label{dn-dk.vs.delta}
\eal
which is actually the Eq.~\eqn{dn.vs.k} that we assumed to interpret Eq.~\eqn{transport-eq} 
as a genuine Boltzmann kinetic equation. \\
Another important derivative of the equilibrium distribution that is required to study the 
response to a temperature gradient is 
\be
\fract{\partial n^0_{\ep\bk a}}{\partial T} = \bigg(\!\!-\!
\fract{\partial f(\ep)}{\partial \ep}\!\bigg)\Bigg[\fract{\ep}{T}\fract{\partial \delta_a(\ep,\bk)}{\partial \ep}
+ \fract{\partial \delta_a(\ep,\bk)}{\partial T}\;\Bigg]\;,\label{BBBB}
\ee
where   
\beal
\fract{\partial \delta_a(\ep,\bk)}{\partial T} = 
\Ima\Bigg[\;G_+(\ep,\bk)\;\fract{\partial \Sigma_+(\ep,\bk)}{\partial T}\;\Bigg]\,.
\label{AAAA}
\eal
Inspection of the perturbative expansion of the self-energy in terms of skeleton diagrams 
leads to the following result
\beal
\fract{\partial \Sigma_{a}(\ep,\bk)}{\partial T} &= 
\fract{1}{V}\sum_{\bk a'}\int d\ep'\, \fract{\partial f(\ep')}{\partial T}\\
&\;
\Gamma^q_{a,a';a',a}(\ep\,\bk,\ep'\,\bk';\ep'\,\bk',\ep\,\bk)\;
\mathcal{A}(\ep',\bk')
\;,
\eal
where $\Gamma^q$ is the $q$-limit of the reducible vertex, which implies that    
\beal
\fract{\partial \Sigma_{+ a}(\ep,\bk)}{\partial T} &= Z_{* }(\ep,\bk)^{-1}\;
\fract{1}{V}\,\sum_{\bk' a'}\,\int d\ep'\, \fract{\partial f(\ep')}{\partial T}\\
&\qquad A^q_{a,b}(\ep\,\bk,\ep'\,\bk')\;
\mathcal{A}_{\text{qp}}(\ep',\bk')
\;.
\eal
In conclusion, through \eqn{AAAA}, we find 
\beal
\fract{\partial \delta_a(\ep,\bk)}{\partial T} 
&= -\mathcal{A}_{\text{qp}\,a}(\ep,\bk)\;
\fract{\pi}{V}\,\sum_{\bk' a'}\,\int d\ep'\, \fract{\partial f(\ep')}{\partial T}\\
&\qquad A^q_{a,a'}(\ep\,\bk,\ep'\,\bk')\;
\mathcal{A}_{\text{qp}}(\ep',\bk')
\;,
\eal
and thus Eq.~\eqn{BBBB} becomes 
\bw
\beal
\fract{\partial n^0_{\ep \bk a}}{\partial T}
&= \bigg(\!\!-\!
\fract{\partial f(\ep)}{\partial \ep}\!\bigg)\,
\mathcal{A}_{\text{qp}}(\ep,\bk)\,\Bigg[\;
\fract{\ep}{T} - \fract{1}{V}\,\sum_{\bk' a'}\,\int d\ep'\, 
\bigg(\!\!-\!
\fract{\partial f(\ep')}{\partial \ep'}\!\bigg)\,\,\fract{\ep'}{T}\,A^q_{a,a'}(\ep\,\bk,\ep'\,\bk')\;
\mathcal{A}_{\text{qp}}(\ep',\bk')\,\Bigg]\;.
\eal
\ew
If we instead consider the derivative 
with respect to $T$ of the local equilibrium Wigner distribution \eqn{local-equilibrium}, its expression greatly simplifies by  
making use of Eq.~\eqn{A.vs.f} that relates $A^q$ to the Landau $f$-parameters, leading to 
\beal
\fract{\partial\, \overline{n}^{\,0}_{\ep \bk a}}{\partial T} &= 
-\fract{\partial f(\ep)}{\partial \ep}\;\fract{\ep}{T}\;\mathcal{A}_\text{qp}(\ep,\bk)
\,.\label{important: dn/dT}
\eal

\bibliographystyle{apsrev}

\begin{thebibliography}{27}
\bibitem[{not({\natexlab{a}})}]{nota-Gamma}
\bibinfo{note}{We mention that the derivation relies also on the assumption
  that the irreducible vertex in the particle-hole channel with energy transfer
  $\omega$ and momentum transfer $\bq$ is analytic at $\omega=|\bq|=0$. {T}his is
  however a very plausible assumption, and it is hard to imagine a situation
  where it may not hold.}
\bibitem[{not({\natexlab{b}})}]{nota-phase-shift}
\bibinfo{note}{Such link implies, e.g., that a genuine quasiparticle
  corresponds to a resonance, i.e., a phase shift of $\pi/2$, that becomes
  narrower and narrower approaching the Fermi surface, or, in the case of the
  self-energy {E}q.~\eqn{singular-Sigma}, to two resonances at opposite sides of
  the chemical potential that interfere destructively at $\ep=0$.}
\end{thebibliography}

\end{document}